\documentclass[oldversion]{aa}

\usepackage[tbtags,fleqn]{amsmath}
\usepackage{amsfonts}
\usepackage{pstricks, pst-plot}
\usepackage{graphicx}
\usepackage{natbib}

\newcommand{\diff}{\mathrm d}

\newcommand{\erf}{\mathrm{erf}}

\newcommand{\mincir}{\raise
  -2.truept\hbox{\rlap{\hbox{$\sim$}}\raise5.truept \hbox{$<$}\ }}
\newcommand{\magcir}{\raise
  -2.truept\hbox{\rlap{\hbox{$\sim$}}\raise5.truept \hbox{$>$}\ }}


\begin{document}

\title{Larson's third law and the universality of molecular cloud
  structure} 
\author{Marco Lombardi\inst{1,2}, Jo\~ao Alves\inst{3}, and Charles
  J. Lada\inst{4}}
\titlerunning{Larson's 3rd law and the universality of molecular cloud
  structure}
\authorrunning{M. Lombardi, J. Alves, C.J. Lada}
\offprints{M. Lombardi}
\mail{mlombard@eso.org}
\institute{%
  University of Milan, Department of Physics, via Celoria 16, I-20133
  Milan, Italy
  \and 
  European Southern Observatory, Karl-Schwarzschild-Stra\ss e 2, 
  D-85748 Garching bei M\"unchen, Germany 
  \and 
  University of Vienna, T\"urkenschanzstrasse 17, 1180 Vienna, Austria
  \and
  Harvard-Smithsonian Center for Astrophysics, Mail Stop 42, 60 Garden
  Street, Cambridge, MA 02138}
\date{Received ***date***; Accepted ***date***} 

\abstract{%
  Larson (1981) first noted a scaling relation between masses and
  sizes in molecular clouds that implies that these objects have
  approximately constant column densities.  This original claim, based
  upon millimeter observations of carbon monoxide lines, has been
  challenged by many theorists, arguing that the apparent constant
  column density observed is merely the result of the limited dynamic
  range of observations, and that in reality clouds have column
  density variations over two orders of magnitudes.  In this letter we
  investigate a set of nearby molecular clouds with near-infrared
  excess methods, which guarantee very large dynamic ranges and robust
  column density measurements, to test the validity of Larson's third
  law.  We verify that different clouds have almost identical average
  column densities \textit{above a given extinction threshold\/}; this
  holds regardless of the extinction threshold, but the actual
  average surface mass density is a function of the specific threshold
  used.  We show that a second version of Larson's third law,
  involving the mass-radius relation for single clouds and cores, does
  not hold in our sample, indicating that individual clouds are not
  objects that can be described by constant column density.  Our
  results instead indicate that molecular clouds are characterized by
  a universal structure.  Finally we point out that this universal
  structure can be linked to the log-normal nature of cloud column
  density distributions.  }
\maketitle

%

\defcitealias{2001A&A...377.1023L}{Paper~0}
\defcitealias{2006A&A...454..781L}{Paper~I}
\defcitealias{Lombardi08a}{Paper~II}
\defcitealias{Lombardi09a}{Paper~III}

\section{Introduction}
\label{sec:introduction}

It has long been recognized that star formation is inextricably linked
to the molecular clouds where the process is taking place, and
therefore it is important to study the structure of these objects.
One of the first attempts in this direction has been carried out by
\citet{1981MNRAS.194..809L}.  In his seminal work, Larson used
molecular line data available from earlier studies (mostly millimeter
data of nearly objects) and showed that molecular clouds obey three
scaling relations: (1) a power-law relationship between the length $L$
of the cloud and its velocity dispersion $\sigma_\mathrm{v}$, with
$\sigma_\mathrm{v} \propto L^{0.38}$; (2) approximate virial
equilibrium, with $2 G M / \sigma_\mathrm{v}^2 L \simeq 1$; and (3) a
relationship between the density $n$ of the cloud and its length,
with $n \propto L^{-1.1}$.  Larson's third law, which is the main
focus of this letter, implies that molecular clouds have approximately
constant column densities $\Sigma$, since $\Sigma \sim n L \propto
 L^{-0.1}$.

 Since their formulation, Larson's laws have been the subject of
 several observational and theoretical studies.  From the
 observational point of view, \citet{1987ApJ...319..730S} presented
 ${}^{12}$CO data for a 273 nearby clouds from the University of
 Massachusetts-Stony Brook (UMSB) Galactic Plane Survey
 \citep{1986ApJS...60....1S}.  They found a size-line width
 relationship with a power index (0.5) steeper than the one derived by
 \citet{1981MNRAS.194..809L}.  Additionally, in agreement with
 Larson's third law, they found that the molecular gas surface density
 is approximately constant for all clouds with $\Sigma(\mathrm{H}_2) =
 170 \mbox{ M}_\odot \mbox{ pc}^{-2}$.  Recently, the same sample of
 clouds has been reanalysed by \citet{2009ApJ...699.1092H} using data
 from the Boston University-FCRAO Galactic Ring Survey
 \citep{2006ApJS..163..145J}.  The use of ${}^{13}$CO ($J = 1–0$)
 emission instead of ${}^{12}$CO ensures that a large fraction of the
 data are optically thin; additionally, the data used have a much
 higher angular sampling and spectral resolution.
 \citet{2009ApJ...699.1092H} confirmed Larson's third law with a
 relative scatter (approximately a factor 3) similar to previous
 studies.  However, surprisingly they found a median mass surface
 density of molecular hydrogen for this sample of $42 \mbox{ M}_\odot
 \mbox{ pc}^{−2}$, thus significantly smaller than the one derived by
 \citet{1987ApJ...319..730S}.

On the theoretical side, there have been many attempts to explain
Larson's laws using numerical simulations.  In many cases, the
validity of Larson’s relations, and especially of the third law, has
been questioned \citep{1989A&A...225..517K, 1990ASSL..162..151S,
  1997ApJ...474..292V, 2002ApJ...570..734B, 2006MNRAS.372..443B}.  In
particular, it has been suggested that this law is merely the result
of the limited dynamic range of observations, and that in reality mass
surface densities of molecular clouds span at least two orders of
magnitude.

In this letter, we re-examine the validity of Larson's third law using
extinction as a tracer of molecular gas \citep{1994ApJ...429..694L}.
The use of this tracer, in combination with advanced techniques
\citep{2001A&A...377.1023L, 2009A&A...493..735L}, allows us to probe
clouds over a large dynamical range (typically more than two order of
magnitudes in extinction); additionally, the column density
measurements use a simple tracer, dust, which is not plagued by the
uncertainties affecting millimeter observations of gas and dust (e.g.,
deviations from local thermodynamic equilibrium, opacity variations,
chemical evolution, small-scale structure, depletion of molecules,
unknown emissivity properties of the dust, unknown dust temperature).

The results of this study are twofold: first, we verify that Larson's
law of constant column density holds with a very small scatter on a
set of nearby clouds investigated using \textsc{Nicer}
\citep{2001A&A...377.1023L} and \textsc{Nicest}
\citep{2009A&A...493..735L}; second, we show that the same law,
applied within a single cloud (using different extinction thresholds)
as $M \propto L^2$ does not hold.  Additionally, we argue that the
first version of Larson's third law implies a universal physical
structure for molecular clouds, which we identify in their log-normal
distributions for the projected gas density.

Larson's third law, in its original formulation, links the average
density $\bigl\langle n(\mathrm{H}_2) \bigr\rangle$ of clouds with
their size $L$: $\bigl\langle n(\mathrm{H}_2) \bigr\rangle = 3\,400
\mbox{ cm}^{-3} (L / 1 \mbox{ pc})^\alpha$, with $\alpha = -1.10$.
Here $L$ is defined as the maximum observed linear extent of the
cloud, and $\bigl\langle n(\mathrm{H}_2) \bigr\rangle$ is the average
density of a sphere of diameter $L$ and total mass $M$ identical to
the cloud (typically estimated from ${}^{13}\mbox{CO}$ data).
Larson's data were more heterogeneous and included different clouds
studied at different contours of integrated intensity, which resulted
in a scatter of approximately one order of magnitude about the assumed
relation; as we will see, our data suggest instead that Larson's law
holds with a scatter below $15\%$.  The fact that $\alpha \simeq -1$
implies that the cloud projected column density, $\bigl\langle
n(\mathrm{H}_2) \bigr\rangle L \propto L^{-0.1}$, is approximately
constant.  Larson discussed a few possible explanations for this:
one-dimensional shock compressions, optical depth natural selection
effects, and observational biases owing to the limited dynamic range of
the ${}^{13}\mbox{CO}$ data.

\section{An extinction measurement of Larson's law}
\label{sec:an-extinct-meas}

\subsection{Definitions}
\label{sec:definitions}

We consider first (Sect.~\ref{sec:larsons-third-law-1}) the following
version of Larson's third law.  Since we have at our disposal complete
extinction maps, we can consider the area $S$ of a cloud \textit{above
  a given extinction threshold\/} $A_0$ (throughout this letter,
unless otherwise noted, we will refer to extinction measurements in
the $K$ band, $A_K$, and drop everywhere the index $K$).  We then
define the cloud size implicitly from $S = \pi (L/2)^2$ (or the cloud
radius as $R = L/2$).  Similarly, we can consider the cloud mass $M$
above the same extinction threshold.

We will also briefly investigate the mass vs. radius relationship for
each individual cloud, and verify whether we recover Larson's
prediction $M(R) \propto R^2$ (Sect.~\ref{sec:larsons-third-law-2}).
Note that the two versions of Larson's third law (different clouds
above a fixed extinction threshold, or same cloud at various
extinction thresholds) are clearly linked, but are not equivalent, in
the sense that only one of the two might hold.  Note also
\citet{1981MNRAS.194..809L} de-facto studied different clouds at
different thresholds, and therefore used a mixture of both versions
considered separately here.

Throughout this letter we will treat molecular complexes as single
objects, and we will not split unconnected regions belonging to the
same complex.  Since typically a cloud will have many clumps with
relatively high column densities, this procedure avoids the
``creation'' of new clouds when the extinction threshold $A_0$ is
increased.  This procedure is justified because our objects are mainly
well defined regions, relatively far from the galactic plane, and with
no or little contamination from other clouds.

\subsection{Data analysis}
\label{sec:data-analysis}

The data used here are extinction maps obtained from the point source
catalog of the Two Micron All Sky Survey
\citep[2MASS;][]{1994ExA.....3...65K}.  Data for the various complexes
have been reduced using \textsc{Nicer} \citep{2001A&A...377.1023L} and
\textsc{Nicest} \citep{2009A&A...493..735L} and following the
prescriptions adopted in previous works (see
\citealp{2006A&A...454..781L, 2008A&A...489..143L,
  2010A&A...512A..67L}).  The complexes considered are nearby
molecular clouds, and therefore we are able to well resolve most cores
with the 2MASS data; the same clouds have been used in
\cite{Charlie2010}.  Extinction measurements are converted into
surface mass densities using
\begin{equation}
  \label{eq:1}
  \Sigma = \mu m_\mathrm{p} \beta_K A_K \; ,
\end{equation}
where $\mu$ is the mean molecular weight, $\beta_K \equiv
[N(\mathsc{Hi}) + 2 N(\mathrm{H}_2)] / A_K \simeq 1.67 \times 10^{22}
\mbox{ cm}^{-2} \mbox{ mag}^{-1}$ is the gas-to-dust ratio
\citep{1979ARA&A..17...73S,1955ApJ...121..559L,1978ApJ...224..132B},
and $m_\mathrm{p}$ is the proton mass.  With a standard gas
composition (63\% hydrogen, 36\% helium, and 1\% dust) we have $\mu
\simeq 1.37$ and therefore $\Sigma / A_K \simeq 183 \mbox{ M}_\odot
\mbox{ pc}^{-2} \mbox{ mag}^{-1}$.

\begin{figure}
  \centering
  \includegraphics[width=\hsize]{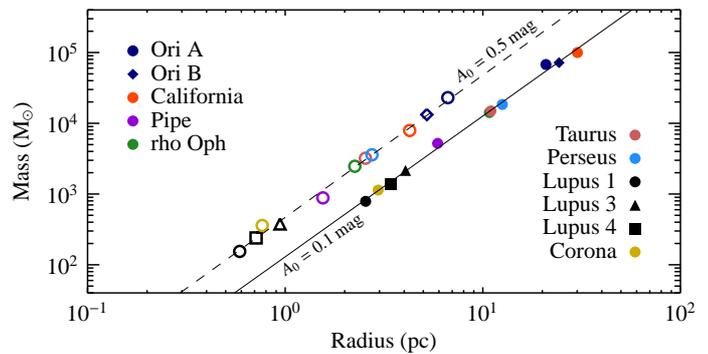}
  \caption{Cloud masses above extinction thresholds of $A_0 = 0.1
    \mbox{ mag}$ (filled symbols) and $A_0 = 0.5 \mbox{ mag}$ (open
    symbols) as a function of their size.  The two line shows the best
    constant surface density fits, which correspond to $\Sigma = 41
    \mbox{ M}_\odot \mbox{ pc}^{-2}$ and $\Sigma = 149 \mbox{ M}_\odot
    \mbox{ pc}^{-2}$ respectively.}
  \label{fig:1}
\end{figure}

\begin{table}[b]
  \centering
  \begin{tabular}{ccccc}
    Threshold $A_0$ & $a$ & $\gamma$ & Scatter & $c$ \\
    (mag) & ($\mbox{M}_\odot \mbox{ pc}^{-\gamma}$) & & (percent) & \\
    \hline
    $0.1$ & $\phantom{0}41.2$ & $1.99$ & $11\%$ & $2.25$ \\
    $0.2$ & $\phantom{0}73.1$ & $1.96$ & $12\%$ & $2.00$ \\
    $0.5$ & $          149.0$ & $2.01$ & $14\%$ & $1.63$ \\
    $1.0$ & $          264.2$ & $2.06$ & $12\%$ & $1.44$ \\
    $1.5$ & $          379.8$ & $2.07$ & $14\%$ & $1.38$ \\
  \end{tabular}
  \caption{Best power-law fits $M = a \pi R^\gamma$ for various
    extinction thresholds.  Note that because $\gamma \simeq 2$ in all
    cases, the quantity $a$ can be interpreted as the average mass
    column density of the cloud above the corresponding extinction
    threshold.  The last two columns show the standard
    deviation of the cloud column densities divided by their
    average (relative scatter) and the ratio between the average
    column densities and the minimum column density set by
    the extinction threshold ($c$).} 
  \label{tab:1}
\end{table}

\subsection{Larson's third law for a constant extinction threshold}
\label{sec:larsons-third-law-1}

Figure~\ref{fig:1} shows the amount of mass different clouds have
above extinction thresholds of $A_K = 0.1 \mbox{ mag}$ and $A_K = 0.5
\mbox{ mag}$ as a function of the cloud ``radii'' (defined according
to Sect.~\ref{sec:definitions}), together with the best power-law
fit.  As apparent from this plot, all clouds follow exquisitely well a
Larson-type relationship, with $M \propto R^2$, and have therefore
very similar projected mass densities \textit{at each extinction
  threshold}.  This result is also quantitatively shown in
Table~\ref{tab:1}, where we report the best-fit power-laws for
the mass vs.\ radius relation at different extinction thresholds.  The
exceptionally small scatter observed in Fig.~\ref{fig:1} is also
confirmed by the results shown in Table~\ref{tab:1}: \textit{at all
  extinctions considered, data follow the best-fit power-laws with
  relative standard deviations always below $15\%$.}

Table~\ref{tab:1} also show the dimensionless factor $c$ obtained from
the best quadratic fit $M = c \mu m_\mathrm{p} \beta_K A_0 \pi R^2$ in
terms of the constants appearing in Eq.~\eqref{eq:1}.  Hence, $c =
\langle A_K \rangle / A_0 \ge 1$, and the fact that $c \sim 2$ with a
very small relative scatter among different clouds indicates that all
these objects have a very similar physical structure.

\subsection{Larson's third law for single clouds}
\label{sec:larsons-third-law-2}

\begin{figure}
  \centering
  \includegraphics[width=\hsize]{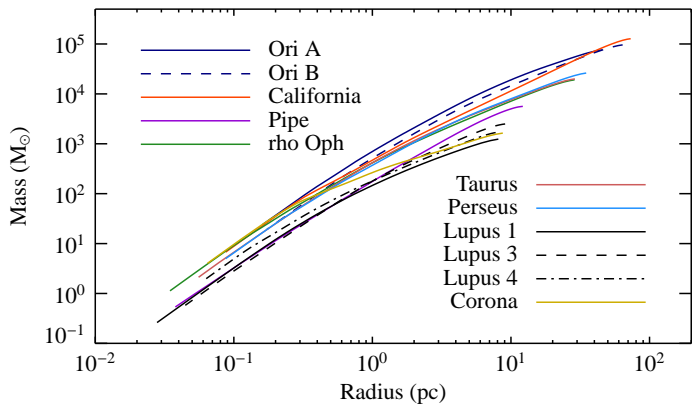}
  \caption{Mass vs.\ radius relationship; both quantities are defined
    as indicated in Sect.~\ref{sec:definitions}.}
  \label{fig:2}
\end{figure}

Figure~\ref{fig:2} shows the second version of Larson's third law
considered here, i.e.\ the mass vs.\ radius relationship.  As apparent
from this figure, the tracks for the various clouds have similar
trends, but span a relatively large range of masses.  In the range $R
\in [0.1,1] \mbox{ pc}$ we can fit a power-law of the form $M(R) = 380
\mbox{ M}_\odot \, (R / \mathrm{pc})^{1.6}$, a result that compares
well with the one obtained by \citet{2010ApJ...716..433K}, $M(R) = 400
\mbox{ M}_\odot \, (R / \mathrm{pc})^{1.7}$.  Different clouds have
quite similar exponents (the standard deviation of the power-law index
is $\sim 0.18$), but rather different masses (the best-fit scale
parameter for the mass ranges from $150$ to $710 \mbox{ M}_\odot$).
Note, however, that since the power-law index is significantly
different from two, errors on the assumed distances of the clouds
would affect the scale parameter for the mass.  

From this analysis we conclude that Larson's third law is not an
accurate description of the mass vs.\ radius relationship for single
clouds.  Specifically, at larger scales all clouds show a flattening
of the curves and deviates significantly from a power-law, while at
smaller scales clouds follow power-laws, but with an exponent
significantly different than two.

\subsection{Cloud physical structure}
\label{sec:cloud-phys-struct}

\begin{figure}
  \centering
  \includegraphics[width=\hsize]{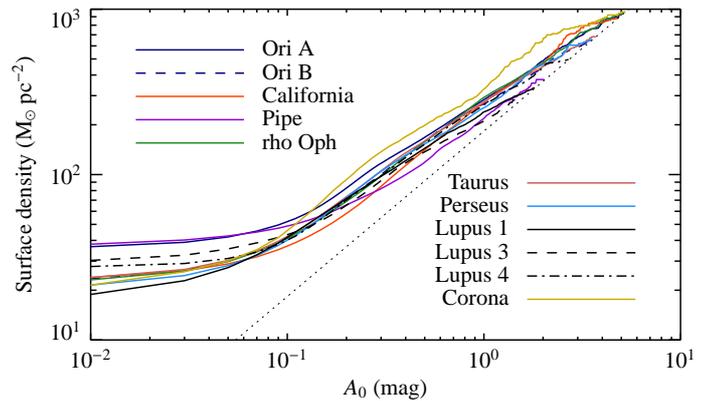}
  \caption{Cloud mass surface density above an extinction threshold as
    a function of the threshold, in logarithmic scale.  The dotted
    line shows the relationship between the cloud column density in
    $\mbox{M}_\odot \mbox{ pc}^{-2}$ and the extinction in the $K$
    band.}
  \label{fig:3}
\end{figure}

As mentioned earlier, that an ensemble of clouds satisfies Larson's
third law at different extinction thresholds suggests that clouds have
a universal physical structure.

In order to investigate this point better, we consider in
Fig.~\ref{fig:3} the average column density of cloud material above a
given extinction threshold, as a function of the extinction threshold.
Figure~\ref{fig:3} indicates a remarkable uniformity among the various
clouds: they all show a relatively flat plateau up to $\sim 0.1 \mbox{
  mag}$, and then a constant rise up to $2$--$5 \mbox{ mag}$.  In the
range $A_0 \in [0.1, 1] \mbox{ mag}$, the curves for all clouds are
confined within a relatively narrow region.  In this extinction range
we can fit a simple power-law to the data plotted in Fig.~\ref{fig:3},
obtaining $\Sigma = 265 \mbox{ M}_\odot \mbox{ pc}^{-2} \, (A_0 /
\mbox{mag})^{0.8}$.  Note that an error analysis of the data points in
Fig.~\ref{fig:3} at $A_0 < 0.05 \mbox{ mag}$ shows that they are
significant, because the large number of independent measurements
contributing to these data make the statistical errors negligible, and
because the flatness of the plateau at low extinction values makes
them robust with respect to systematic errors (such as offsets in the
\textsc{Nicer} maps due to extinction in the control field).

\section{Theoretical interpretation}
\label{sec:theor-interpr}

\begin{figure}
  \centering
  \includegraphics[width=\hsize]{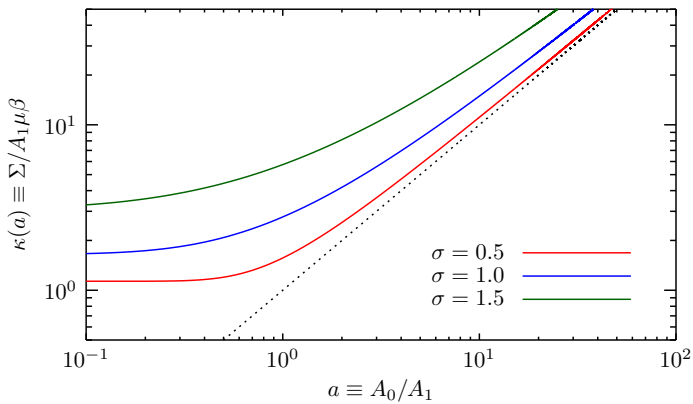}
  \caption{Dimensionless column density $\kappa \equiv \Sigma / A_1
    \mu m_\mathrm{p} \beta$, defined in Eq.~\eqref{eq:6}, as a
    function of the dimensionless column density $a \equiv A_0 / A_1$.
    The dotted line shows the asymptotic limit $\kappa(a) \simeq a$.
    Note the similarity with the curves shown in Fig.~\ref{fig:3}.}
  \label{fig:4}
\end{figure}

The results presented above indicates that clouds have similar
structures.  Observationally (see, e.g., \citealp{2008A&A...489..143L,
  2009A&A...508L..35K, 2010A&A...512A..67L, 2010MNRAS.tmp..812F}),
many clouds show a log-normal distribution at low extinctions:
\begin{equation}
  \label{eq:2}
  p_A(A) = \frac{1}{\sqrt{2 \pi} \sigma A} 
  \exp \left[ -\frac{(\ln A - \ln A_1)^2}{2 \sigma^2} \right] \; ,
\end{equation}
where $A_1$ and $\sigma$ are two positive parameters.  A tail at high
extinctions, present in many clouds, is generally associated with the
effects of gravitational instability.  The log-normality of $p_A(A)$
is often linked with supersonic turbulence, although recent results
show that this is also a common feature of very different classes of
cloud models \citep{2010arXiv1006.2826T}.

Interestingly, we can express the mass and the area of a cloud above
an extinction threshold as simple integrals of $p_A(A)$.  Given a
cloud of total area $S_\mathrm{tot}$, the area and mass above a given
extinction threshold $A_0$ are
\begin{align}
  \label{eq:3}
  S(A_0) & {} = S_\mathrm{tot} \int_{A_0}^\infty
  p_A(A) \, \diff A \; , \\
  \label{eq:4}
  M(A_0) & {} = S_\mathrm{tot} \mu m_\mathrm{p} \beta \int_{A_0}^\infty A
  p_A(A) \, \diff A \; .
\end{align}
In particular, if we consider the log-normal distribution of
Eq.~\eqref{eq:2}, we obtain for the column density above $A_0$
\begin{equation}
  \label{eq:5}
  \Sigma(A_0) \equiv
  \frac{M(A_0)}{S(A_0)} = A_1 \mu m_\mathrm{p} \beta
  \kappa(A_0 / A_1) \; ,
\end{equation}
where $\kappa$ is a \textit{dimensionless\/} quantity defined as
\begin{equation}
  \label{eq:6}
  \kappa(a) = \exp \left( \frac{\sigma^2}{2} \right)
  \frac{1 - \erf\bigl[ ( \ln a - \sigma^2 ) / \sqrt{2} \sigma \bigr]}%
  {1 - \erf\bigl[ \ln a / \sqrt{2} \sigma \bigr]} \; .
\end{equation}
We plot in Fig.~\ref{fig:4} the function $\kappa(a)$ in a log-log
scale for three values of the parameter $\sigma$.  A comparison of
Fig.~\ref{fig:4} with Fig.~\ref{fig:3} shows that the log-normal model
is able to capture the main characteristics, except the ``collapse''
of curves onto the dotted line in Fig.~\ref{fig:3} (most likely due to
limited dynamic range of observations).  Furthermore, in order to
obtain the narrow bundle of curves in Fig.~\ref{fig:3}, the parameters
$A_1$ and $\sigma$ for the various clouds must span a limited range.
Typical relative scatters of $A_1$ and $\sigma$ are of the order of
$57\%$ and $19\%$ (cf.\ Tab.~\ref{tab:2}).  Additionally, an
analytical calculation shows that the particular form of the
log-normal distribution further suppresses these scatters
(respectively by factors between 8 and 4), so that the final expected
relative standard deviation in $\Sigma$ of the order of $14\%$, in
agreement with the data presented in Table~\ref{tab:1}.

\begin{table}
  \tiny
  \begin{minipage}[t]{0.48\hsize}
    \begin{tabular}{lcc}
      Cloud & $A_1$ & $\sigma$ \\ 
      \hline
      Lupus V & 0.15 & 0.42 \\
      Coalsack & 0.38 & 0.28 \\
      Taurus & 0.18 & 0.49 \\
      Lupus I & 0.08 & 0.43 \\
      Ophiuchus & 0.16 & 0.48 \\
      Serpens & 0.33 & 0.51 \\
      Cha I & 0.11 & 0.35 \\
      Cha II & 0.12 & 0.35 \\
      Lupus III & 0.14 & 0.35 \\
      Cor A & 0.10 & 0.44 \\
      LDN 1228 & 0.12 & 0.32 \\
      Pipe & 0.42 & 0.29
    \end{tabular}
  \end{minipage}
  \hfill
  \begin{minipage}{0.48\hsize}
    \begin{tabular}{lcc}
      Cloud & $A_1$ & $\sigma$ \\ 
      \hline
      LDN 134 & 0.10 & 0.39 \\
      LDN 204 & 0.14 & 0.41 \\
      LDN 1333 & 0.12 & 0.38 \\
      LDN 1719 & 0.15 & 0.50 \\
      Musca & 0.08 & 0.45 \\
      Cha 3 & 0.12 & 0.46 \\
      Ori A GMC & 0.13 & 0.50 \\
      Perseus & 0.13 & 0.48 \\
      Ori B GMC & 0.11 & 0.49 \\
      Cepheus & 0.16 & 0.59 \\
      California & 0.14 & 0.51 \\
      & &
    \end{tabular}
  \end{minipage}
  \caption{Log-normal fit parameters for various clouds.  Data are
    from \citet{2009A&A...508L..35K} and converted into $K$-band
    extinction parameters using a standard reddening 
    law \citep{1985ApJ...288..618R}.}
  \label{tab:2}
\end{table}

\section{Summary}
\label{sec:summary}

\begin{enumerate}
\item Using near-infrared extinction maps of a set of nearby clouds we
  tested Larson's third law for molecular clouds, the constancy of
  average mass surface densities above a given extinction threshold.
  We verified this scaling law to a relatively high degree of
  precision.  We found a very small ($<15\%$) relative scatter for the
  measured column densities independent of the adopted extinction
  thresholds over a very large range, from $A_K = 0.1 \mbox{ mag}$ to
  $A_K = 1.5 \mbox{ mag}$.  Additionally, we found the value of the
  average mass surface density to be a function of the adopted
  extinction threshold.
\item We verified that Larson's third law does not hold when
  considering the mass-radius relation within single clouds.  In the
  range $R \in [0.1, 1] \mbox{ pc}$ we find that the mass scales as
  $M(R) \propto R^{1.6}$, and is therefore significantly shallower
  than what was predicted by Larson; at larger radii, the relation
  appears to flatten even more.
\item We interpreted these results, and in particular item~1 above, as
  the effects of a universal physical structure shared among the
  different clouds.  We showed that this universal structure is
  represented by a uniformity in the cloud density distributions.  We
  found that a log-normal model is able to account for this
  uniformity, provided that the log-normal parameters are restricted
  to relatively narrow ranges.  This suggests that Larson's third law
  might be a consequence of this special property of cloud structure.
\end{enumerate}

\acknowledgements We thank the referee for helping us to significantly
improve this paper.

\bibliographystyle{aa} 
\bibliography{../dark-refs}

\end{document}